# ON THE RELATIONSHIP BETWEEN SOFTWARE COMPLEXITY AND SECURITY

Mamdouh Alenezi and Mohammad Zarour

College of Computer and Information Sciences,
Prince Sultan University, Riyadh, Saudi Arabia

## ABSTRACT

*This work aims at discussing the complexity aspect of software while demonstrating its relationship with security. Complexity is an essential part of software; however, numerous studies indicate that they increase the vulnerability of the software systems and introduce bugs in the program. Many developers face difficulty when trying to understand the complex components of software. Complexity in software increases when objects in the software are used to design a more complex object while creating a hierarchical complexity in the system. However, it is necessary for the developers to strive for minimum complexity, as increased complexity introduces security risks in the software, which can cause severe monetary and reputational damage to a government or a private organization. It even causes bodily harm to human beings with various examples found in previous years where security breaches led to severe consequences. Hence it is vital to maintain low complexity and simple design of structure. Various developers tend to introduce deliberate complexities in the system so that they do not have to write the same program twice; however, it is getting problematic for the software organizations as the demands of security are continually increasing.*

## 1. INTRODUCTION

Extensive researches have been conducted around the factor of software complexity in the field of software engineering [8, 1]. As software complexity is the result of implementation details, design decisions, and algorithm choices, it can be used to predict various characteristics of software quality, e.g. maintainability [16, 1]. An example can be taken from the case of highly complex modules that are found to be more vulnerable to faults [25]. Hence, it is highly imperative to monitor and control any software in order to successfully complete the project. It was showed by the research conducted in the field of software engineering that the presence of documentation in a software development project assures the quality of the code of complex functions and the design [1].

Many developers face difficulty in understanding the complicated structures of software units. What makes this problem more difficult is that the complexity level can be varying even in the solution of the same problem, depending on the variation of the implementations [1]. The software also continuously evolves according to user requirement and for this, the software has to be updated on a regular basis which consequently increases the complexity of the software [1]. Whereas, in order to ensure a smooth update process for additional features and bug fixes, it is highly important to keep the complexity of the system to the minimum level [1]. The evolution of the complexity of a closed-source system was investigated by Manduchi and Taliercio [17] and it was found that approximately all the measures relevant to complexity increased at different rates. Another study conducted by MacCormack et. al [16] demonstrated that restructuring the design of Mozilla decreased the complexity based on Lehman's law, which states that until precautionary measures are not implemented, the complexity of the system will continue to increase. The question that remains constant is whether is it true for every case that adding features to the system will always increase its complexity? In order to gain in-depth insight, the present research thoroughly discusses the concept of software complexity while assessing the relationship between security and software complexity.

## 2. RELATED WORK

### 2.1. Software Complexity

In the 10th edition of the Software Engineering book of Ian Sommerville [23], he stated that the critical issue found in modern engineering of software is the management of complexity.

Complexity is the degree to which the design or implementation of a system or its component is difficult to verify and understand [15]. As stated by Zuse [28], the actual meaning of complexity can be said to be the difficulty found in understanding, maintaining, and modifying the code. Whereas, according to Fenton and Bieman [10], the resources spent on the development and maintenance of a solution for any task can be viewed as the software complexity. Correspondingly, software complexity is viewed by Basili [5] as a measure of resource allocation through a system or a user while interacting with a component of software for performing a task. While these studies maintain the understanding that difficulty in comprehension arises from complexity, they do not further explore complexity. Briand, et al. [6] states that the definition of complexity should be in the form of a fundamental part of software instead of a problem; however, in information theory, the complexity is defined by Kolmogorov [14] as the minimum possible length for a system description in some languages. Nevertheless, it is not always easy to find the minimum possible length of a system; thereby, the sophistication of this definition can be found in its focus on the essence of the measurement and complexity of complexity.

Hence, software complexity can be seen as an indicator of a lack of project quality [1]. An empirical investigation shows that complexity provides an advantage in understanding the mechanism behind the evolution and management of complexity. More specifically, size and complexity are two factors that make accurate predictions about external attributes such as flexibility and maintainability [20]. McGraw stated that complexity is regarded as the enemy of security [31]. Every software has a different level and structure of complexity [27]. Complexity in the structure of software that can be found because of the cluster of program elements includes:

1) **Architectural complexity** arises because of a high number of interconnected subsystems.
2) **Cyclomatic complexity** arises because of a high number of loops and if statements. It has been found that cyclomatic complexity has a direct influence on security [29].
3) **Cognitive complexity** can be discovered by observing the behavior of users finding a solution to a particular task or problem. Its main focus is on the understandability aspect of the code.
   - Lack of understanding can lead to causing mistakes, which will consequently introduce vulnerability in the system.
   - Extensive deliberation about the working of a feature.
   - When the method is subjective, it is significant.

Complexity in the inputs can also be a cause of introducing vulnerabilities in the system and huge security risks, such as application to the operating system or pages to web browsers.

## 2.2. No Vulnerabilities vs. No obvious Vulnerabilities

Software complexity is the main reason for the vulnerabilities found in the system [13]. Multiple software practitioners have created a hypothesis that the systems, that are not designed and/or maintained efficiently and effectively, are more susceptible to vulnerabilities. In contrast to this, various studies have assessed the relationship between complexity and vulnerability and concluded that it cannot be generalized [22, 19]. A Microsoft study conducted by Morrison et al. [19], stated that there is no significant connection between complexity and vulnerability. The study further recommended the use of more security-specific metrics in the models for vulnerability prediction.

Software complexity is said to be one of the main properties of a software system [13]. It is also known as the source of most of the bugs in the software. In the empirical study of [7], 52 releases of Mozilla Firefox that had been developed over the course of 4 years, were used to correlate complexity metrics and security vulnerabilities. It was observed that the correlation between complexity metrics and the vulnerabilities in Mozilla Firefox was positive (i.e. slightly more than 0.5).

In another exploratory study, a dataset was collected from 2875 security patches that contained the metrics and vulnerabilities associated with all the functions, files, and classes of 5750 different versions of projects that are widely used, i.e. Mozilla, Linux Kernel, Xen Hypervisor, Glibc, and HTTPD [2]. Multiple statistical tests and correlations were conducted in this study in order to discover the connection between vulnerabilities and metrics. The results of the study showed that the complexity metrics are able to consistently display some characteristics found in codes that make them susceptible to vulnerabilities. Furthermore, the study also shows that complexity metrics result in presenting higher values in the files that are vulnerable. Conclusively, the study stated that across most modules, the metrics tend to be higher in the code that is vulnerable as compared to code that reported no vulnerabilities. Complexity can occur in the development phase of the software. Such as, in the requirement phase, if the proper decomposition of requirements is not executed, they will become more complex and difficult to understand which would consequently lead to issues in the designing and development phases. Similarly, if the proposed architectural design of the software is complex, the developers can face problems while developing complex architecture. Additionally, if in the development phase, the developer writes a code that is complex, it will be difficult for other developers to understand it and cause issues in its maintenance later on. Moreover, the testing phase can also face a similar situations, such that if the testing scenarios and test cases are complex, then their execution, monitoring, and maintenance can get difficult in the future. The main purpose of the software development models is to break down all the activities of the development process to reduce the probability of complexity.

When the functions in the code are complex, software complexity becomes an expected byproduct. When the system contains multiple interfaces and complex set of requirements, the software systems can grow increasingly complex. It usually results in the cost of maintenance for the software systems and portfolios to go over the budget and increase the risks for enhancements. Without proper maintenance, the software complexity will only continue to increase in the delivered projects and make the applications cumbersome and bloated. Any software is relatively complex, and it is probably simplicity at the core of the software which enables the complex systems to be less burdened because of complications [12].

## 2.3. Sources of Vulnerability

It seems that critical vulnerabilities are released on a weekly basis, which then burdens the vendors and the system administrator to remediate. Hence, what is the source of these vulnerabilities? National Vulnerability Database shows a simple search that more than 3,300 vulnerabilities were released in the past 3 months only that are specific to particular applications. The most popular example of egregious vulnerability is Heartbleed, which affects approximately half of the web servers over the internet. Another question that arises at this point is, why having so many vulnerabilities and at such frequency? The main reason can be attributed to the fact that vulnerability is an emergent property of software caused by code quality, trusted data inputs, and complexity.

## 2.4. Reason for the Growing Number of Disclosed Vulnerabilities

An increasing number of disclosed vulnerabilities are due to different reasons.

1. Deployment of a growing number of novel technological services and products is enabled by modern innovations and technology that is continuously evolving. The success of cloud computing, social media, mobile devices, the Internet of Things (IoT), and mobile applications in the market have created opportunities for new innovative businesses, but with relatively low secure-coding skills and maturity in software development.

2. As innovation and competition in the field of technology continue to grow, the number of hardware and software products is increasing. Decreased time-to-market has become a popular demand and has forced the vendors to release products and services at a fast pace, which leads to fewer resources spent on quality assurance and testing.

3. The novel software products have increased complexity because of the demand for integration, interconnectivity, and platform compatibility. Around 80–90% of applications are built using software components that are open-source. This increases the probability of vulnerabilities in a system even more. According to recent research [24], every 1 out of 18 open-source software components that were downloaded in 2017 consisted of a known security vulnerability.

4. The number of people and organizations that are involved in the research and exploration of vulnerabilities is increasing. A recent study [21] suggested that this group is responsible for the growing number of vulnerabilities being disclosed. Research reward programs have been established by major vendors in order to mobilize this group for testing different products for a re-compensating reward. Some examples include Microsoft Security Research Centre and Google Vulnerability Reward Program (VRP).

5. As the cyber-crime is becoming more beneficial in monetary aspects, the growing number of threats is also a part of vulnerability research. Various vulnerabilities are because of cybercriminals searching for new tools with which they can support their attacks.

6. As cybersecurity is becoming a priority, various nations are also involved in the vulnerability research so that lawful intervention can be supported in IT-systems.

7. In 2015, it was reported by [26] that new regulations, which require the presence of backdoors in hardware and software were adopted by a Nation-State. This is known as vulnerability-by-design where backdoors are deliberately left in the software by the vendors for diverse purposes such as malicious intentions, surveillance programs or hidden commercial agendas.

8. Outdated system architectures are also found in the released products for various reasons.

9. Spectre is found to be the recently researched vulnerabilities that have impacted a huge variety of microprocessors that have been launched from 1995 till now. ENISA has recently published an InfoNote [11] to thoroughly demonstrate these vulnerabilities.

10. As there is a lack of adequate legislation and regulation, which protect the consumer rights against the defects and vulnerabilities in software, the industry is not effectively motivated to invest in the security of the product and/or assume the responsibility for the damages.

It can be seen from the literature about software and its security that two factors are always together, i.e. building software that is secure with controlled operationality and managed risks for deployed applications [18]. Although these two factors seem to be similar but are actually very different. In the first factor, the focus is on offering an opportunity for the user to write code, conduct few tests, and develop the application following the rules of SDLC. However, the second factor is associated with providing space where a developer can manage risks related to the deployed application [3]. In this manner, security vulnerabilities of an application may introduce severe security issues in the system if they are not managed or controlled at an early stage. Software insecurities can be found in a system for a number of reasons and related security issues are as follows:

1. Software, specifically the ones connected to the internet, are facing an increased number of sources of insecurities.
2. Software complexity.
3. Assumptions made by the development team for the executed input of external units are inaccurate and inadequate.
4. Lack of premeditated connections between the components of the software, inclusive of those provided by third-party vendors.
5. Appropriate use of morality over improper logical clarity relevant to code behavior.
6. Insufficient validation of input.
7. Exception and error handling

McGraw stated that software behavior and functionality under malicious attack comes under the topic of software security [31], which is a part of software engineering. Security experts have discussed 3 factors, which can be held responsible for the devastation of the logical singularity of predictable outcomes. These three factors are greatly impacting the growth and evaluation process of software development processes. According to a study, with the rise of these components in the development process of application, the presence of malicious code will be easily identified. The three factors are complexity, connectivity, and extensibility. Since the present research is focused on the aspect of complexity, only this factor is discussed in detail.

As the growth of the information system is rapidly increasing, the major cause of security flaws can be found in the form of complexity. Common reasons for complexity are the increment in code size and storage capacity. As the development complexity increases along with the lines of code, it becomes more difficult to detect any defects in the system [30]. In a large system, it is almost impossible to avoid any bugs. With increased connectivity, code complexity, sophisticated attacks, and easier accessibility to attack tools have led to an overwhelming and continuously increasing number of security breaches that can cause severe damage to several software organizations. While complexity alone does not make things incomprehensible, its excessiveness can increase the difficulty in understanding things. Nevertheless, the complexity of the system should never exceed a defined limit, as it will consequently lose the effectiveness of the system in terms of user-friendliness, financial viability, and other prospects while decreasing the overall market value of the product. Consequently, it is imperative to keep complexity to a certain limit to effectively manage software security, see Figure 1. Hence, complexity can be a major cause of security breaches [32].

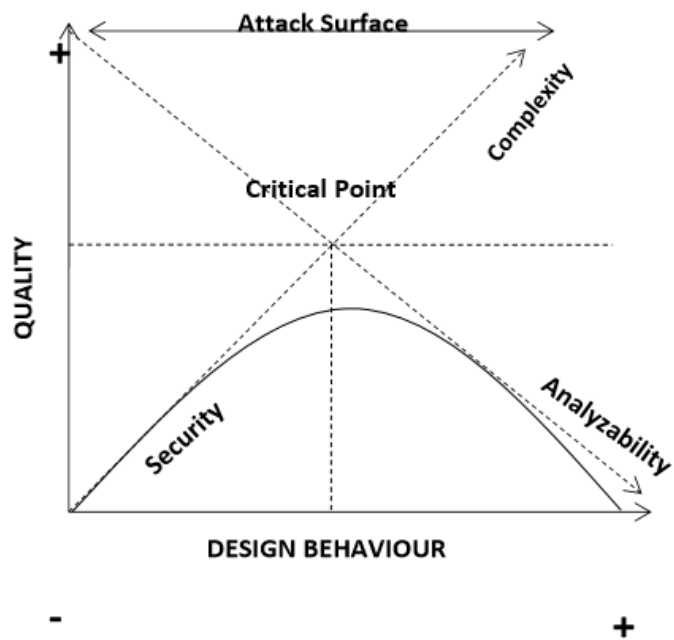

Figure 1. Security vs Complexity

Increased complexity makes it easier for an application to completely breakdown. High complexity results in an over-budgeted, deficient and delayed product, opposite to the requirements of any project. Thus, the question arises on how to deal with this issue? One answer is to thoroughly study how the complexity in other systems is managed. There is even complexity in a personal computer, such as a CPU consist of main memory, an arithmetic/logic unit (ALU) and a bus with which other devices are connected. Each component can be divided further. ALU can be broken down into registers and random control logic, which is further split into NAND gates, inventors, etc. and it continues. As indicated by Brooks, complexity should be considered as a significant property of software. The complexity can be derived from various elements, for instance, the problem domain complexity, possible flexibility through software, the difficulty in the management of software development processes, and the issues faced in the characterization of discrete system behaviors.

Consequently, the observations that are found to be common in all the complex systems are:

1. A complex system is made up of subsystems that are interrelated and form a network of subsystems; hence, they create a complexity hierarchy.
2. While exploring the nature of the primitive components in a complex system, deducing which components are primitive is done arbitrarily and is largely dependent on the person observing the system.
3. Linkages that are intra components are commonly stronger than the ones that are inter components and this affects the separation of high-frequency dynamics in the components.
4. The internal structure of the components is also involved which consists of low-frequency dynamics; this feature facilitates interaction between the components. Often, the implementation of complex systems is done with an economy of expressions, which further demonstrates that complex systems consist of common patterns.

A simple system that works further elaborates that designing complex systems from scratch is barely conducive and it is impossible to work. In a simple system, a platform is provided for the processes to start over from the beginning. With the evolution of the system, the complex objects are used to build other more complex systems. It is not possible to correctly craft primitive objects. First, this should be used accordingly and then should be enhanced with the actual behavior of the system.

## 2.5. Security Engineering: An Inclusive Approach

In this era, the software organizations are facing pressure from the users of their products to focus more on the security aspect, while developing the software. Today, a security breach of software can lead to severe consequences. The vulnerabilities found in the software programs can cause severe loss to the government and private organizations, alike in the monetary and reputation aspects and have also become a possible danger to human life. As information systems become applicable in almost all scenarios, security becomes an even more crucial aspect of software engineering.

Consequently, the reasons for faulty software should be highlighted by software professionals to allow other people to understand the roots of security problems. Today's designers are under the obligation to take security very seriously as defects in the system are a major cause of problems of security. Security designers should design strong mechanisms, which would address the faults in the design and bugs in implementation at the time of development. As this study contributes to the knowledge of the field, it is significant in terms of highlighting the importance of:

- Creating a better understanding of design and architecture information of the software system, which can consequently enable easier comprehension of the process and maintenance.
- Discovering and minimizing the vulnerabilities in the early stages of the software development process will help in making the end product secure.
- Evaluating the security of software and calculating the cost estimation of the project would help in the estimation and planning of new activities.

Security is theorized as multidimensional and greasy. It can be perceived in various ways along with the risks associated with business goals. The main objective is to discover the issues relevant to the software at an early stage. The proposed security measures can be used to mitigate the security issues in risk management, knowledge, and best practices. With the risk management framework, security risks can be identified, synthesized, mitigated and validated in the context of business value. Security is a crucial issue and it is imperative for everyone to learn about it to provide optimal service and support. Security is a multidimensional attribute through which a precisely specified and perspective framework, a better process and method/roadmap are ensured. Forgoing explanations make it evident that assessment techniques or a framework of security are critical; however, they should be of prescriptive nature and can be used easily to minimize vulnerability and measure design security.

Every type of security is a concern for the functional and non-functional activities of the organization, also known as Business Process Security, which is key for any venture. Various hardware and software security tools are specified to provide security trials during the organization's online activities. However, strong security systems are still required to deal with the threats in data transactions. Currently, numerous organizations are involved in the development of software tools or resources, which would allow them to deal with the pressure of handling security effectively and provide protection to their businesses against malicious attacks and risk factors.

Risk factors are being defined by experts in terms of threats, vulnerabilities, and consequences. However, the developers are unaware of the security measures that they need to take in the development processes. If the contributions made by the experts in the field of security are to be used in an effective manner, it is essential to maintain a causal relationship between attacks, threats, intrusion, vulnerability, failure, damage to resources, risk, error, security indexing, and prioritization. An empirical study conducted by Assal and Chiasson [4] observed that simple design principles are not followed by many practitioners and they deliberate introducing complexity in the system to avoid rewriting a code. As the design principles are not adopted, a complex system riddled with flaws will be generated that would increase the security problems [9]. Furthermore, evaluating security can become difficult because of complex design and lacking readability [4].

## 3. CONCLUSIONS

This work discusses the issue of software complexity and how it leads to vulnerable software. After researching various studies, it can be deduced that the higher the complexity of a software system is, the more chances of security flaws are. Complexity creates difficulty in the ability of developers to understand the software, and hence, affect the maintainability of the system along with difficulties in introducing updates to it. While numerous studies indicated that the main reason for vulnerabilities in a software system is because of increased complexity, some studies also stated that the relation between vulnerabilities and complexity is vague; thus, it cannot be generalized. Complexity in the system occurs in the development process of the software, usually at the development phase, designing phase, and/or requirement phase. Various security experts have studied and contributed to the security measures that are necessary for the development of secure software; however, developers either remain ignorant of it or introduce deliberate complexities in the system in order to avoid rewriting the codes. This is problematic for organizations because customers demand not only quick delivery of the products, but also demand strong product security. Products with no or minimum security do not survive long, irrespective of what they might be offering to the consumers.